# Spinor superalgebra: Towards a theory for higher spin particles

A. D. Alhaidari

*Physics Department, King Fahd University of Petroleum & Minerals,*
*Dhahran 31261, Saudi Arabia*
E-mail: haidari@mailaps.org

We define a superalgebra $S^2(N/2)$ as a $Z_2$ graded algebra of dimension $2N+3$, where $N$ is a positive, odd integer. The even component is a three-dimensional abelian subalgebra, while the odd component is made up of two $N$-dimensional, mutually conjugate algebras. For $N = 1$, two of the three even elements become identical, resulting in a four-dimensional superalgebra which is the graded extension of the SO(2,1) Lie algebra that has recently been introduced in the solution of the Dirac equation for spin $1/2$. Realization of the elements of $S^2(N/2)$ is given in terms of differential matrix operators acting on an $N+1$ dimensional space that could support a representation of the Lorentz space-time symmetry group for spin $N/2$. The $N = 3$ case results in a 4×4 matrix wave equation, which is linear and of first order in the space-time derivatives. We show that the "canonical" form of the Dirac Hamiltonian is an element of this superalgebra.

## I. INTRODUCTION

Several theories for free and interacting fields and particles of higher spin have been proposed over the years since the original work of Dirac [1], Fierz and Pauli [2], Rarita and Schwinger [3], Bargmann and Wigner [4], and others. Most are plagued with one or more of the known difficulties associated with existing theories of spin larger than one. For a recent review of these developments and citations to the relevant literature, one may consult [5] and references therein. Consequently, one may choose not to disagree with the view that, in this case, "unconventional" wisdom might be given an opportunity to attempt a solution for this persistent and difficult problem, often referred to as the "high spin crisis". In this article, we build on our recent experience with solutions of the relativistic wave equation for spin $1/2$ [6-9] by formulating a theory, equivalent to that of a massive relativistic particle of spin $N/2$, from an algebraic point of view. As an application, we consider the one-dimensional problem of a spinor with non-minimal coupling to the relativistic potential. We specialize to the case where the "even component" of the potential vanishes, resulting in a "canonical form" of the theory. Other non-canonical problems that belong to the same class as this reference (canonical) problem could be solved by mapping the reference problem into others, using for example the "extended point canonical transformation (XPCT)" introduced in [6]. The preceding terminology in quotes was borrowed from Rfs. [6-9].

The theory presented here (in its canonical form) is constructed using a differential matrix operator realization of a $2N+3$ dimensional $Z_2$ graded algebra, which we refer to as the "Spinor Superalgebra $S^2(N/2)$." We start by giving a brief account of the recent development that led to the findings which included this superalgebra for $N = 1$, and which consequently provided the motivation for the present attempt to investigate higher spins.

We recently presented an effective approach for solving the three-dimensional Dirac equation for spherically symmetric local interactions in a series of articles [6-9]. The first step in this development started with the realization that the nonrelativistic Coulomb, Oscillator, and S-wave Morse problems belong to the same class of potentials which carry a representation of SO(2,1) Lie algebra. Therefore, the fact that the relativistic versions of the first two problems (Dirac-Coulomb and Dirac-Oscillator) were solved exactly makes the solution of the third, in principle, feasible. Indeed, the relativistic S-wave Dirac-Morse problem has been formulated and solved exactly [7]. The bound state energy spectrum and spinor wave functions were obtained. Taking the nonrelativistic limit reproduces the familiar S-wave Schrödinger-Morse problem. Subsequently, the same approach was applied successfully in obtaining solutions for the relativistic extension of yet another class of shape-invariant potentials [8]. These included the Dirac-Scarf, Dirac-Rosen-Morse I & II, Dirac-Pöschl-Teller, and Dirac-Eckart potentials. Furthermore, using the same formalism, quasi exactly solvable systems with rest mass energies were obtained for a class of power-law relativistic potentials [9]. Quite recently, Guo Jian-You *et al* succeeded in constructing solutions for the relativistic Dirac-Woods-Saxon [10] and Dirac-Hulthén [11,12] problems using the same approach. Exact $L^2$ series solutions for all energies – discrete (for bound states) and continuous (for scattering states) – were obtained for several relativistic problems, including the Dirac-Coulomb [13] and Dirac-Morse [14]. These were written in terms of the Meixner-Pollaczek and the continuous dual Hahn orthogonal polynomials, respectively. Moreover, the solution of the Dirac equation with position-dependent mass in the Coulomb field has been obtained [15]. The relativistic two-point Green's function was constructed for the Dirac-Morse [16] and Dirac-Oscillator [17] problems. In the fourth article of the series in our program of searching for exact solutions to the Dirac equation, we found a special graded extension of SO(2,1) Lie algebra [6]. Realization of this superalgebra in terms of 2×2 matrices of differential operators acting in the two-component spinor space was constructed. The linear span of this graded algebra gives the "canonical form" of the radial Dirac Hamiltonian. It turned out that the Dirac-Oscillator class, which also includes the Dirac-Coulomb and Dirac-Morse, carries a representation of this supersymmetry. Another such class includes the Dirac-Scarf, Dirac-Pöschl-Teller, Dirac-Hulthén, etc. [18]. In the following section, a brief summary will be given of this approach, whose aim is to separate the variables, such that the two coupled, first order differential equations resulting from the Dirac equation generate a Schrödinger-like equation. This makes the solution of the relativistic problem easily attainable by simple and direct correspondence with well-known, exactly solvable, nonrelativistic problems. The summary will also be useful for those who are not familiar with the subject, and will also assist in giving a natural transition to the introduction of the superalgebra, which is the essential ingredient in the present work. In section III, we extend these findings (the superalgebra associated with spin $1/2$) to spin $3/2$ and introduce $S^2(3/2)$. This will be followed, in section IV, by defining the generalized spinor superalgebra $S^2(N/2)$. In Section V, we conclude with a short discussion about the three-dimensional problem with spherical symmetry and, as an example, obtain the relativistic energy spectrum for a spin $3/2$ particle in the Dirac-Oscillator potential.

## II. SPIN $1/2$ REPRESENTATION OF $S^2(1/2)$ :
## GRADED EXTENSION OF SO(2,1)

To give a simple and clear presentation of our findings, which were obtained while searching for solutions to the relativistic wave equation for spin $1/2$ particles, we consider the one-dimensional problem. In the atomic units $\hbar = m = 1$, the most general Dirac Hamiltonian for a two-component spinor includes couplings to the time-independent vector potential $(A_0, A_1) = (V, \hbar^{-1}U)$, scalar potential $S$, and pseudo-scalar potential $W$. It reads as follows [19]:

$$H = \begin{pmatrix} 1 + \hbar^2\left[V(x) + S(x)\right] & \hbar\left[U(x) + iW(x) - i\dfrac{d}{dx}\right] \\ \hbar\left[U(x) - iW(x) - i\dfrac{d}{dx}\right] & -1 + \hbar^2\left[V(x) - S(x)\right] \end{pmatrix}, \qquad (2.1)$$

where $\hbar$ is the Compton wavelength $\hbar/mc = c^{-1}$. $H$ is measured in units of the rest mass, $mc^2 = \hbar^{-2}$. Gauge invariance of the vector potential coupling could be used to eliminate the contribution of the off diagonal term $\hbar U$ in the Hamiltonian (2.1). Moreover, writing the two component spinor as $\psi = \begin{pmatrix} if^+(x) \\ f^-(x) \end{pmatrix}$ results in the following $2 \times 2$ matrix wave equation:

$$\begin{pmatrix} 1 + \hbar^2(V + S) - \varepsilon & \hbar\left(W - \dfrac{d}{dx}\right) \\ \hbar\left(W + \dfrac{d}{dx}\right) & -1 + \hbar^2(V - S) - \varepsilon \end{pmatrix} \begin{pmatrix} f^+ \\ f^- \end{pmatrix} = 0, \qquad (2.2)$$

where $f^{\pm}(x)$ are the two spinor components which are real and square-integrable functions, and $\varepsilon$ is the relativistic energy, which is measured in units of $mc^2$. $V \pm S$ and $W$ are real functions referred to as the "even" and "odd" components of the relativistic potential, respectively. Equation (2.2) results in two coupled, first order differential equations for the two spinor components. Eliminating one component in favor of the other gives a second order differential equation. This will not be Schrödinger-like (i.e., it contains first order derivatives) unless $S = \pm V$. That is, unless one of the diagonal elements in the Hamiltonian matrix is a constant. To obtain a Schrödinger-like equation in the general case we proceed as follows: a global unitary transformation $\mathcal{U}(\eta) = \exp(\frac{1}{2}\hbar\eta\sigma_2)$ is applied to the Dirac equation (2.2), where $\eta$ is a real parameter and $\sigma_2$ is the $2 \times 2$ matrix $\begin{pmatrix} 0 & -i \\ i & 0 \end{pmatrix}$. The Schrödinger-like requirement relates the potential functions by the linear constraint $cS + \frac{s}{\hbar}W = \pm V$, where $s = \sin(\hbar\eta)$, $c = \cos(\hbar\eta)$ and $-\frac{\pi}{2} \le \hbar\eta \le +\frac{\pi}{2}$. This results in a Hamiltonian that will be written in terms of only two arbitrary potential functions. The unitary transformation together with the potential constraint map Eq. (2.2) into the following one, which we choose to write in terms of $V$ and $W$ as:

$$\begin{pmatrix} c - \varepsilon + (1\pm1)\lambdabar^2 V & \lambdabar\left[ -\dfrac{s}{\lambdabar} + \dfrac{1}{c}W \mp \dfrac{s}{c}\lambdabar V - \dfrac{d}{dx} \right] \\ \lambdabar\left[ -\dfrac{s}{\lambdabar} + \dfrac{1}{c}W \mp \dfrac{s}{c}\lambdabar V + \dfrac{d}{dx} \right] & -c - \varepsilon + (1\mp1)\lambdabar^2 V \end{pmatrix} \begin{pmatrix} \phi^+ \\ \phi^- \end{pmatrix} = 0, \quad (2.3)$$

where

$$\begin{pmatrix} \phi^+ \\ \phi^- \end{pmatrix} = \mathcal{U}\psi = \begin{pmatrix} \cos\dfrac{\lambdabar\eta}{2} & \sin\dfrac{\lambdabar\eta}{2} \\ -\sin\dfrac{\lambdabar\eta}{2} & \cos\dfrac{\lambdabar\eta}{2} \end{pmatrix} \begin{pmatrix} f^+ \\ f^- \end{pmatrix}. \tag{2.4}$$

This gives the following equation for one spinor component in terms of the other:

$$\phi^\mp(x) = \frac{\lambdabar}{c\pm\varepsilon}\left[ \mp\frac{s}{\lambdabar} \pm \frac{1}{c}W(x) - \frac{s}{c}\lambdabar V(x) + \frac{d}{dx} \right]\phi^\pm(x), \tag{2.5}$$

where $\varepsilon \neq \mp c$. This is known as the "kinetic balance" relation. On the other hand, the resulting Schrödinger-like wave equation for the two spinor components reads as follows:

$$\left[ -\frac{d^2}{dx^2} + G^2 \mp \frac{dG}{dx} + 2\left(\varepsilon \pm \frac{1}{c}\right)V - 2\frac{s}{c\lambdabar}W - \frac{\varepsilon^2-1}{\lambdabar^2} \right]\phi^\pm(x) = 0, \tag{2.6}$$

where $G = \frac{1}{c}W \mp \frac{s}{c}\lambdabar V$, and the top/bottom sign corresponds to the choice of sign in the potential constraint $cS + \frac{s}{\lambdabar}W = \pm V$. In all relativistic problems that have been successfully tackled so far, Eq. (2.6) is solved by correspondence with well-known, exactly solvable, nonrelativistic problems [7-12]. This correspondence results in a parameter map relating the relativistic to the nonrelativistic problem. Using this map and the known solutions (energy spectrum and wavefunctions) of the nonrelativistic problem, one can easily and directly obtain the relativistic energy spectrum and spinor wave functions. For example, when $S = 0$, three of the many problems that have already been solved are:

1) Dirac-Morse: $V(x) = -Ae^{-\omega x}$, where $A$ and $\omega$ are real and $\omega > 0$

$$\left[ -\frac{d^2}{dx^2} + \left(\frac{c}{s}\lambdabar A\right)^2 e^{-2\omega x} - A\left(2\varepsilon - \frac{c}{s}\lambdabar\omega\right)e^{-\omega x} - \frac{\varepsilon^2-1}{\lambdabar^2} \right]\phi^\pm(x) = 0. \quad (2.7)$$

2) Dirac-Hulthén: $V(x) = -A/(e^{\omega x}-1)$

$$\left[ -\frac{d^2}{dx^2} + \frac{\frac{c}{s}\lambdabar A(\frac{c}{s}\lambdabar A + \omega)}{(e^{\omega x}-1)^2} - A\frac{2\varepsilon + \frac{c}{s}\lambdabar\omega}{e^{\omega x}-1} - \frac{\varepsilon^2-1}{\lambdabar^2} \right]\phi^\pm(x) = 0. \tag{2.8}$$

3) Dirac-Oscillator: $V(x) = 0$, $W(x) = Ax$

$$\left[ -\frac{d^2}{dx^2} + A^2x^2 \mp A - \frac{\varepsilon^2-1}{\lambdabar^2} \right]\phi^\pm(x) = 0. \tag{2.9}$$

In fact, the relativistic extension of any known dynamical relationship in the nonrelativistic theory could easily be obtained by the correspondence map. The Green's function, which has prime significance in the calculation of physical processes, is such an example [16,17].

An alternative approach is to map the known solution of a given reference problem into the intended target problem that belongs to the same class as the reference problem, using for example the Darboux transformation or point canonical transformation. This is due to the fact that all analytically solvable problems fall within distinct classes that carry some representation of a given symmetry group. Examples are found in the class of exactly solvable problems whose potentials are referred to as the "shape invariant" potentials [20].

Next, we intend to show that the class of problems with $V = S = 0$ (the "canonical" class) carries a representation of a supersymmetry algebra, which is a special and nontrivial graded extension of SO(2,1) Lie algebra, and which we designate as $S^2(1/2)$. Specifically, we will establish that the Dirac operator in Eq. (2.2) with $V = S = 0$, which we refer to as the "canonical" Dirac operator, is an element of the $S^2(1/2)$ superalgebra. This class corresponds to problems with the superpotentials $W^2 \pm W'$, because the wave equation (2.6) in this case reads as follows:

$$\left[ -\frac{d^2}{dx^2} + W^2 \mp W' - \frac{\varepsilon^2 - 1}{\lambda^2} \right] \phi^{\pm} = 0 \qquad (2.10)$$

The Dirac-Oscillator, where $W \sim x$, is such an example. The Dirac-Scarf and Dirac-Pöschl-Teller potentials are also two examples, among others, of elements in this class [18,21].

SO(2,1) is the three-dimensional Lie algebra with basis elements satisfying the commutation relations $[L_3, L_\pm] = \pm L_\pm$ and $[L_+, L_-] = -L_3$. It is a very useful and highly important algebra in various physical applications, and in the solution of many three-parameter problems. It has been studied extensively in the literature as a potential algebra and spectrum-generating algebra for several problems [22]. We define a graded extension of this algebra as the four-dimensional superalgebra $S^2(1/2)$, with two odd elements $L_\pm$ and two even elements, $\hat{L}$ and $L_3$, satisfying commutation / anti-commutation relations [6]:

$$[L_3, L_\pm] = \pm L_\pm, \ \{L_+, L_-\} = \hat{L}, \ [\hat{L}, L_3] = [\hat{L}, L_\pm] = 0 , \qquad (2.11)$$

where $L_\pm^\dagger = L_\mp$, which implies hermiticity of the even operators (i.e., $\hat{L}^\dagger = \hat{L}$ and $L_3^\dagger = L_3$). These relations also show that $\hat{L}$ forms the center of this superalgebra, since it commutes with all of its elements. We construct the following realization of the elements of this graded Lie algebra as 2×2 matrices of differential operators acting in the two-component spinor space:

$$L_\pm = \sigma_\pm \left[ W(x) \mp \frac{d}{dx} \right], \ L_3 = \frac{1}{2} \sigma_3 , \qquad (2.12)$$

where $\sigma_+ = \left( \begin{smallmatrix} 0 & 1 \\ 0 & 0 \end{smallmatrix} \right)$, $\sigma_- = \left( \begin{smallmatrix} 0 & 0 \\ 1 & 0 \end{smallmatrix} \right)$, $\sigma_3 = \left( \begin{smallmatrix} 1 & 0 \\ 0 & -1 \end{smallmatrix} \right)$ and $W(x)$ is a real differentiable function. Using the anticommutation relation in (2.11) we obtain:

$$\hat{L} = \begin{pmatrix} -\dfrac{d^2}{dx^2} + W^2 - W' & 0 \\ 0 & -\dfrac{d^2}{dx^2} + W^2 + W' \end{pmatrix} , \qquad (2.13)$$

where $W' = dW/dx$. The odd operators $L_\pm$ are the raising and lowering operators in the two-component spinor space. They are first order in the derivatives, whereas the even operators are zero and second order. $L_3$ is the spin operator and parity operator. If $Q \in S^2(1/2)$, so that it carries a representation of this supersymmetry, then $Q$ could be expanded as a linear combination of these four basis elements as follows:

$$Q = \lambda_+ L_+ + \lambda_- L_- + \lambda_3 L_3 + \hat{\lambda}\hat{L}, \tag{2.14}$$

where the $\lambda$'s are constant parameters. Requiring that this operator be hermitian and a first order differential results in $\hat{\lambda}_- = \hat{\lambda}_+^*$, $\hat{\lambda}_3^* = \hat{\lambda}_3$, and $\hat{\lambda} = 0$, which yields:

$$Q = \begin{pmatrix} 1 & \hbar\left[W(x) - \dfrac{d}{dx}\right] \\ \hbar\left[W(x) + \dfrac{d}{dx}\right] & -1 \end{pmatrix}, \tag{2.15}$$

where $\hbar \equiv 2\lambda_+/\lambda_3$ and $Q$ has been renormalized with the nonzero "mass factor" $\lambda_3/2$. Since $\hat{L}$ is the center of the superalgebra and commutes with $Q$, then a two-component representation, $\chi$, could be found such that $Q|\chi\rangle \sim |\chi\rangle$ and $\hat{L}|\chi\rangle \sim |\chi\rangle$. Using this observation in the comparison of Eq. (2.15) with Eq. (2.2) and Eq. (2.13) with Eq. (2.6) shows that the class of relativistic problems for spin $1/2$ particles with $V = S = 0$ carries a representation of the superalgebra $S^2(1/2)$, i.e. $\chi = \begin{pmatrix} f^+ \\ f^- \end{pmatrix}$. On the other hand, solutions of problems where $V \neq 0$ or $S \neq 0$ could be obtained by one of two methods: (i) using a global unitary transformation similar to that which mapped Eq. (2.2) into Eq. (2.3), or (ii) using, for example, a Darboux transformation or point canonical transformation (PCT) [23] to map the solution of a canonical (reference) problem into the intended target problem that belongs to the same class. In [6,18] an "extended point canonical transformation (XPCT)" was defined, which is the relativistic version of PCT. Just like PCT in the nonrelativistic theory, XPCT maps the solution of the reference problem (i.e., the canonical problem where $V = S = 0$) into those of other relativistic problems (where $V \neq 0$ or $S \neq 0$) that belong to the same class as the reference problem. In fact, XPCT maps the Dirac-Oscillator into the Dirac-Coulomb and Dirac-Morse problems. In the following section, the above developments for $S^2(1/2)$ will be extended to $N = 3$ and $S^2(3/2)$ superalgebra. The connection to the theory of spin $3/2$ will be exploited.

## III. $S^2(3/2)$ AND SPIN $3/2$ REPRESENTATION

A massive spin $3/2$ particle has four degrees of freedom, which we represent by the wave function components $\left\{\phi_k^\pm\right\}_{k=0,1}$ corresponding to positive and negative spin projections, respectively. The four component spinor $\psi$, associated spin operator $L_3$, and parity operator $L_0$ could be represented as follows:

$$\psi = \begin{pmatrix} \phi_1^+ \\ \phi_0^+ \\ \phi_0^- \\ \phi_1^- \end{pmatrix}, \quad L_3 = \frac{1}{2}\begin{pmatrix} 3 & 0 & 0 & 0 \\ 0 & 1 & 0 & 0 \\ 0 & 0 & -1 & 0 \\ 0 & 0 & 0 & -3 \end{pmatrix}, \quad L_0 = \frac{1}{2}\begin{pmatrix} 1 & 0 & 0 & 0 \\ 0 & 1 & 0 & 0 \\ 0 & 0 & -1 & 0 \\ 0 & 0 & 0 & -1 \end{pmatrix}. \tag{3.1}$$

A count of the number of degrees of freedom makes it obvious that there exist three independent raising and three independent lowering operators in this four-dimensional spinor space. We designate this set of operators by $\left\{{}^nL_\pm\right\}_{n=1,2,3}$, where $\left({}^nL_\pm\right)^\dagger = {}^nL_\mp$. The most general structure of the matrix representation of these operators in the four-dimensional spinor space is as follows:

$$^1L_+=\begin{pmatrix} 0 & \times & 0 & 0 \\ 0 & 0 & \times & 0 \\ 0 & 0 & 0 & \times \\ 0 & 0 & 0 & 0 \end{pmatrix},\quad {}^2L_+=\begin{pmatrix} 0 & 0 & \times & 0 \\ 0 & 0 & 0 & \times \\ 0 & 0 & 0 & 0 \\ 0 & 0 & 0 & 0 \end{pmatrix},\quad {}^3L_+=\begin{pmatrix} 0 & 0 & 0 & \times \\ 0 & 0 & 0 & 0 \\ 0 & 0 & 0 & 0 \\ 0 & 0 & 0 & 0 \end{pmatrix},\qquad (3.2)$$

where $\times$ stands for 1×1 linear associative operators that carry the dynamical information of the system, being arbitrary at present, but which will be specified shortly. One can easily verify that ONLY with this general structure that these operators satisfy the "spin commutation" relation $[L_3, {}^nL_\pm] = \pm n\,{}^nL_\pm$. On the other hand, only ${}^2L_\pm$ and ${}^3L_\pm$ satisfy the "parity commutation" relation $[L_0, {}^nL_\pm] = \pm\,{}^nL_\pm$. This is due to the fact that these two operators have a well-defined $\pm$ odd parity, while ${}^1L_\pm$ does not. This could simply be understood by dividing each of these 4×4 matrices into four blocks of 2×2 matrices. Then only ${}^1L_\pm$ has both non-vanishing diagonal (even) and off-diagonal (odd) blocks. To provide ${}^1L_\pm$ with a well-defined parity equivalent to its partners ${}^2L_\pm$ and ${}^3L_\pm$ (i.e., $\pm$ odd parity, respectively), we rewrite it as follows:

$$^1L_+=\begin{pmatrix} 0 & 0 & 0 & 0 \\ 0 & 0 & \times & 0 \\ 0 & 0 & 0 & 0 \\ 0 & 0 & 0 & 0 \end{pmatrix}.\qquad (3.3)$$

Now, *without* any further constraints, these operators satisfy the following relations:

$$[L_0, {}^nL_\pm] = \pm\,{}^nL_\pm,\ [L_3, {}^nL_\pm] = \pm n\,{}^nL_\pm,\ [L_0, L_3] = 0,\ \left\{{}^nL_\pm, {}^mL_\pm\right\} = 0,\qquad (3.4a)$$

where $n, m = 1, 2, 3$. One has to investigate the only remaining relation in order to provide a complete definition of the algebra of these operators, which is the anti-commutation of the odd operators, $\left\{{}^nL_+, {}^mL_-\right\}$. This obviously results in an even operator that adds to the even component of the superalgebra. If $n = m$ for all $n$ and $m$, then the result is a special *even* operator which is *hermitian* and *diagonal*. We call this operator $\hat{L}$, and define it as follows:

$$\hat{L} \equiv \sum_{n=1}^{3} \left\{{}^nL_+, {}^nL_-\right\}.\qquad (3.4b)$$

However, if $n \neq m$, one can then attach physical significance only to the sum of all terms $\left\{{}^nL_+, {}^mL_-\right\}$ with $n - m = $ constant, because those are the terms that produce the same spin shift (i.e., equal raising or lowering steps). That is, we only need to be concerned with the sum $\sum_{n=1}^{3-m} \left\{{}^nL_\pm, {}^{n+m}L_\mp\right\}$ for $m = 1$ and 2. When $m = 2$, we automatically obtain the trivial relation that $\left\{{}^1L_\pm, {}^3L_\mp\right\} = 0$. Therefore, the only remaining non-trivial relation is the ($m = 1$) sum: $\left\{{}^1L_\pm, {}^2L_\mp\right\}$ + $\left\{{}^2L_\pm, {}^3L_\mp\right\}$. Actually, these produce the same entries that we had to eliminate from ${}^1L_\mp$ above. Thus, we should require that they vanish, such that:

$$\left\{{}^1L_\pm, {}^2L_\mp\right\} + \left\{{}^2L_\pm, {}^3L_\mp\right\} = 0.\qquad (3.4c)$$

Consequently, the even component of the superalgebra closes, by the addition of $\hat{L}$, into a three-dimensional abelian subalgebra, where:

$$[L_0, \hat{L}] = [L_3, \hat{L}] = 0.\qquad (3.4d)$$

Finally, we only need to calculate the commutation relation $[\hat{L}, {}^nL_\pm]$ to close the whole superalgebra. Assuming that the relativistic theory we intend to develop for spin 3/2 particles involves only a linear first order differential wave operator, then we can show that the superalgebra $S^2(3/2)$ that we are pursuing is nine-dimensional and that it is closed by the following commutation relation:

$$[\hat{L}, {}^nL_\pm] = 0 .$$ (3.4e)

The assumption that has just been made below Eq. (3.4d) allows us to write the odd operators of the superalgebra as ${}^nL_\pm = {}^n\sigma_\pm\left[W_n(x) \mp \dfrac{d}{dx}\right]$, where $W_n(x)$ are real differentiable functions containing the dynamics of the system, and ${}^n\sigma_\pm$ are the following 4×4 matrices:

$$
{}^1\sigma_+ = \begin{pmatrix} 0 & 0 & 0 & 0 \\ 0 & 0 & a & 0 \\ 0 & 0 & 0 & 0 \\ 0 & 0 & 0 & 0 \end{pmatrix}, \;
{}^2\sigma_+ = \begin{pmatrix} 0 & 0 & b & 0 \\ 0 & 0 & 0 & c \\ 0 & 0 & 0 & 0 \\ 0 & 0 & 0 & 0 \end{pmatrix}, \;
{}^3\sigma_+ = \begin{pmatrix} 0 & 0 & 0 & d \\ 0 & 0 & 0 & 0 \\ 0 & 0 & 0 & 0 \\ 0 & 0 & 0 & 0 \end{pmatrix},
$$ (3.5)

where ${}^n\sigma_- = \left({}^n\sigma_+\right)^\dagger$ and the constant parameters $a$, $b$, $c$ and $d$ are generally complex. Relation (3.4c) results in the following requirements:

$$|c| = |b|, \; d = -b\left(\dfrac{a}{c}\right)^*, \text{ and } W_1 = W_2 = W_3 \equiv W$$ (3.6)

Consequently, we obtain the following matrix representation for the even operator $\hat{L}$ :

$$
\hat{L} = \left(|a|^2 + |b|^2\right)\begin{pmatrix} -\frac{d^2}{dx^2} + W^2 - W' & 0 & 0 & 0 \\ 0 & -\frac{d^2}{dx^2} + W^2 - W' & 0 & 0 \\ 0 & 0 & -\frac{d^2}{dx^2} + W^2 + W' & 0 \\ 0 & 0 & 0 & -\frac{d^2}{dx^2} + W^2 + W' \end{pmatrix} .
$$ (3.7)

Using this representation and

$${}^nL_\pm = {}^n\sigma_\pm\left[W(x) \mp \dfrac{d}{dx}\right],$$ (3.8)

one can easily verify relation (3.4e).

The defining relations (3.4a-3.4e) of the superalgebra $S^2(3/2)$ show that the even operator $\hat{L}$ forms the center of the algebra because it commutes with all of its elements. Moreover, since the three-dimensional even subalgebra is abelian and contains $\hat{L}$, then all representations of $S^2(3/2)$ could be parameterized by three numbers corresponding to the eigenvalues of each of these three mutually commuting operators. We refer to such a representation, whose basis elements are the spinor components $\left\{\phi_k^\pm(x, E)\right\}_{k=0,1}$, as $D_\pm^3(E, k)$, where

$$L_0\left|\phi_k^\pm\right\rangle = \pm\left|\phi_k^\pm\right\rangle, \; L_3\left|\phi_k^\pm\right\rangle = \pm\left(k + \tfrac{1}{2}\right)\left|\phi_k^\pm\right\rangle,$$ (3.9a)

$$\frac{1}{|a|^2 + |b|^2}\hat{L}\left|\phi_k^\pm\right\rangle = \left(-\frac{d^2}{dx^2} + W^2 \mp W'\right)\left|\phi_k^\pm\right\rangle = 2E\left|\phi_k^\pm\right\rangle,$$ (3.9b)

where $E$ is a real parameter and $k = 0$ or $1$. In supersymmetric quantum mechanics [24], $W^2 \pm W'$ are two superpartner potentials giving an interpretation to $\hat{L}$ as the associated Hamiltonian and $E$, the energy. A linear operator $Q$ belongs to $S^2(3/2)$ if we could expand it as:

$$Q = \lambda_0 L_0 + \lambda_3 L_3 + \hat{\lambda}\hat{L} + \sum_{n=1}^3\left(\lambda_n^+ \, {}^nL_+ + \lambda_n^- \, {}^nL_-\right),$$ (3.10)

where $\lambda$'s represent constant parameters. Requiring that $Q$ be real and first order in the derivatives gives $\hat{\lambda} = 0$ and $\lambda_n^- = \left(\lambda_n^+\right)^*$, and results in the following two equivalent matrix representations:

$$Q_\pm = \begin{pmatrix} 1+3\rho & 0 & \nu\left(W-\frac{d}{dx}\right) & \mp\varphi\left(W-\frac{d}{dx}\right) \\ 0 & 1+\rho & \mu\left(W-\frac{d}{dx}\right) & \pm\nu\left(W-\frac{d}{dx}\right) \\ \nu\left(W+\frac{d}{dx}\right) & \mu\left(W+\frac{d}{dx}\right) & -1-\rho & 0 \\ \mp\varphi\left(W+\frac{d}{dx}\right) & \pm\nu\left(W+\frac{d}{dx}\right) & 0 & -1-3\rho \end{pmatrix},$$ (3.11)

where $Q_\pm$ has been renormalized with the nonzero "mass factor" $\lambda_0/2$, and the real constants are defined, with the help of the parameters relation in (3.6), as follows:

$$\rho = \lambda_3/\lambda_0, \ \mu = 2a\lambda_1^+/\lambda_0, \ \nu = 2b\lambda_2^+/\lambda_0, \ \varphi = 2a\lambda_3^+/\lambda_0.$$ (3.12)

Since $\hat{L}$ commutes with $Q_\pm$, then the four component spinor $\psi$ in (3.1) is a simultaneous eigenvector for both. That is, we should have $Q_\pm|\psi\rangle \sim |\psi\rangle$ and $\hat{L}|\psi\rangle \sim |\psi\rangle$. Compatibility of these two relations, with the use of Eq. (3.9b), yields $\rho = 0$ and $\varphi = \mu$. Equivalently, $\lambda_3 = 0$ and $\lambda_3^+ = \lambda_1^+$. Thus, we finally obtain the following two-parameter matrix operator, which is first order in the derivatives and belongs to $S^2(3/2)$:

$$Q_\pm = \begin{pmatrix} 1 & 0 & \nu\left(W-\frac{d}{dx}\right) & \mp\mu\left(W-\frac{d}{dx}\right) \\ 0 & 1 & \mu\left(W-\frac{d}{dx}\right) & \pm\nu\left(W-\frac{d}{dx}\right) \\ \nu\left(W+\frac{d}{dx}\right) & \mu\left(W+\frac{d}{dx}\right) & -1 & 0 \\ \mp\mu\left(W+\frac{d}{dx}\right) & \pm\nu\left(W+\frac{d}{dx}\right) & 0 & -1 \end{pmatrix}.$$ (3.13)

Next, we show that the Dirac Hamiltonian for a spin $3/2$ particle with non-minimal coupling to the two component potential $(A_0, A_1) = (V, \lambdabar^{-1}W)$ in its canonical form (where the time component of the potential vanishes, i.e. $V = 0$) is identical to this representation of $Q_\pm$.

In the atomic units $\hbar = m = 1$, the Compton wavelength $\lambdabar = \hbar/mc = 1/c$ is the relativistic parameter, and the one-dimensional Dirac equation for a free particle takes the form $\left(i\gamma^\alpha\partial_\alpha - \lambdabar^{-1}\right)\psi = 0$, where $\psi$ is the spinor wave function. Use is made of the summation convention over repeated indices. That is, $\gamma^\alpha\partial_\alpha = \gamma^0\partial_0 + \gamma^1\partial_1 = \gamma^0\frac{\partial}{c\partial t} + \gamma^1\frac{\partial}{\partial x}$, where $\gamma^0$ and $\gamma^1$ are anticommuting unimodular constant square matrices satisfying the following relation:

$$\left\{\gamma^\alpha, \gamma^\beta\right\} = \gamma^\alpha\gamma^\beta + \gamma^\alpha\gamma^\beta = 2\mathcal{G}^{\alpha\beta},$$ (3.14)

where $\mathcal{G}$ is the space-time metric, which is equal to diag$(+,-)$. Thus, $\gamma_0^2 = 1$ and $\gamma_1^2 = -1$. These matrices are four-dimensional for spin $3/2$. A matrix representation that satisfies relation (3.14) is chosen as follows:

$$\gamma^0 = \begin{pmatrix} 1 & 0 & 0 & 0 \\ 0 & 1 & 0 & 0 \\ 0 & 0 & -1 & 0 \\ 0 & 0 & 0 & -1 \end{pmatrix}, \quad \gamma^1 = \begin{pmatrix} 0 & 0 & \xi & \mp\zeta \\ 0 & 0 & \zeta & \pm\xi \\ -\xi & -\zeta & 0 & 0 \\ \pm\zeta & \mp\xi & 0 & 0 \end{pmatrix}, \tag{3.15}$$

where $\xi = \cos\theta$, $\zeta = \sin\theta$, and the angular parameter $\pi > \theta > 0$. Next, we allow the Dirac spinor to be coupled to the two-component vector potential $A_\alpha = (A_0, A_1)$. Gauge-invariant coupling, which is accomplished by the "minimal" substitution $\partial_\alpha \rightarrow \partial_\alpha + i\lambdabar A_\alpha$, transforms the free Dirac equation into the following form:

$$\left[ i\gamma^\alpha (\partial_\alpha + i\lambdabar A_\alpha) - \lambdabar^{-1} \right] \psi = 0, \tag{3.16}$$

which, when written in detail, reads as follows:

$$i\frac{\partial}{\partial t}\psi = \left( -i\lambdabar^{-1}\gamma^0\gamma^1 \frac{\partial}{\partial x} + \gamma^0\gamma^1 A_1 + A_0 + \lambdabar^{-2}\gamma^0 \right)\psi. \tag{3.17}$$

For a time-independent potential $(A_0, A_1) = (V, \lambdabar^{-1}W)$, this equation gives the following matrix representation of the Dirac Hamiltonian (in units of $mc^2 = 1/\lambdabar^2$):

$$H = \begin{pmatrix} 1 + \lambdabar^2 V & 0 & \xi\lambdabar\left(W - i\frac{d}{dx}\right) & \mp\zeta\lambdabar\left(W - i\frac{d}{dx}\right) \\ 0 & 1 + \lambdabar^2 V & \zeta\lambdabar\left(W - i\frac{d}{dx}\right) & \pm\xi\lambdabar\left(W - i\frac{d}{dx}\right) \\ \xi\lambdabar\left(W - i\frac{d}{dx}\right) & \zeta\lambdabar\left(W - i\frac{d}{dx}\right) & -1 + \lambdabar^2 V & 0 \\ \mp\zeta\lambdabar\left(W - i\frac{d}{dx}\right) & \pm\xi\lambdabar\left(W - i\frac{d}{dx}\right) & 0 & -1 + \lambdabar^2 V \end{pmatrix}. \tag{3.18}$$

Thus, the eigenvalue wave equation reads $(H - \varepsilon)\chi = 0$, where $\varepsilon$ is the relativistic energy, which is real, dimensionless and measured in units of $1/\lambdabar^2$. Equation (3.16) is invariant under the local gauge transformation:

$$A_\alpha \rightarrow A_\alpha + \partial_\alpha \Lambda, \quad \psi \rightarrow e^{-i\lambdabar\Lambda}\psi, \tag{3.19}$$

where $\Lambda(t, x)$ is a real space-time scalar function. Consequently, the off diagonal terms $\lambdabar W$ in the Hamiltonian (3.18) could be eliminated ("gauged away") by a suitable choice of the gauge field $\Lambda(x)$ in the transformation (3.19). However, our choice of coupling will be non-minimal, which is accomplished by the hermiticity-preserving replacement $W \rightarrow \pm iW$, respectively. If we also write $\psi = \left( i\phi_1^+, i\phi_0^+, \phi_0^-, \phi_1^- \right)^\top$, then we obtain the following 4×4 matrix wave equation:

$$\begin{pmatrix} 1 + \lambdabar^2 V - \varepsilon & 0 & \xi\lambdabar\left(W - \frac{d}{dx}\right) & \mp\zeta\lambdabar\left(W - \frac{d}{dx}\right) \\ 0 & 1 + \lambdabar^2 V - \varepsilon & \zeta\lambdabar\left(W - \frac{d}{dx}\right) & \pm\xi\lambdabar\left(W - \frac{d}{dx}\right) \\ \xi\lambdabar\left(W + \frac{d}{dx}\right) & \zeta\lambdabar\left(W + \frac{d}{dx}\right) & -1 + \lambdabar^2 V - \varepsilon & 0 \\ \mp\zeta\lambdabar\left(W + \frac{d}{dx}\right) & \pm\xi\lambdabar\left(W + \frac{d}{dx}\right) & 0 & -1 + \lambdabar^2 V - \varepsilon \end{pmatrix} \begin{pmatrix} \phi_1^+ \\ \phi_0^+ \\ \phi_0^- \\ \phi_1^- \end{pmatrix} = 0. \tag{3.20}$$

Therefore, taking $\zeta = \mu$, $\xi = \nu$, and $V(x) = 0$ makes the canonical Dirac Hamiltonian in (3.20) identical to the $S^2(3/2)$ hermitian operator $Q_\pm$ in (3.13), and results in the following wave equation for the spinor components $\left\{ \phi_k^\pm \right\}_{k=0,1}$:

$$\left[ -\frac{d^2}{dx^2} + W^2 \mp W' - \frac{\varepsilon^2 - 1}{\lambdabar^2} \right] \phi_k^\pm = 0. \tag{3.21}$$

In the following section we generalize our above findings to the definition of the superalgebra $S^2(N/2)$ for any odd, positive integer $N$.

## IV. THE $S^2(N/2)$ SPINOR SUPERALGEBRA

Following Kac and others [25], a superalgebra $\mathscr{G}$ is defined as a $Z_2$ graded algebra $\mathscr{G} = \mathscr{G}_0 + \mathscr{G}_1$ with a product operation $\circ$ satisfying

$$p \circ q = -(-)^{\sigma(p,q)} q \circ p \,, \tag{4.1}$$

where $\sigma(p,q) = \deg(p) \times \deg(q)$ and

$$\deg(p) = m \leftrightarrow p \in \mathscr{G}_m, \ m = 0 \text{ or } 1. \tag{4.2}$$

An element of $\mathscr{G}$ is called even (odd) if it belongs to $\mathscr{G}_0$ ($\mathscr{G}_1$). We call the anti-symmetric product operation $\circ$ which involves an even element the commutator and designate it by $[\ ,\ ]$ while the symmetric operation that involves only odd elements is called the anticommutator, and is designated by $\{\ ,\ \}$. The superalgebra of interest to our present work is defined with the following properties:

1) $\mathscr{G}_1 = \mathscr{G}_1^+ + \mathscr{G}_1^-$, where $\dim(\mathscr{G}_1^+) = \dim(\mathscr{G}_1^-)$.

2) There exists a one-to-one invertible map (called conjugation): $\mathscr{G}_1^{\pm} \to \mathscr{G}_1^{\mp}$. In other words, for each element $^nL_+ \in \mathscr{G}_1^+$ there corresponds by conjugation an element $^nL_- \in \mathscr{G}_1^-$, where $n = 1,2,..,\dim(\mathscr{G}_1^{\pm})$.

3) $\mathscr{G}_0$, $\mathscr{G}_1^+$, and $\mathscr{G}_1^-$ are abelian subalgebras.

4) $\deg(p \circ q) = \deg(p) + \deg(q)$ if and only if $\sigma(p,q) = 0$. Stated alternatively, $p \circ q \in \mathscr{G}_m$ iff $\deg(p) + \deg(q) = m = 0$ or 1.

5) If $\sigma(p,q) = 1$ (i.e., $p \in \mathscr{G}_1^{\pm}$ and $q \in \mathscr{G}_1^{\mp}$), then only $\left( \sum_{p,q}^{*} p \circ q \right) \in \mathscr{G}_0$, where the sum $\sum^{*}$ is defined in the equivalent statement: $\left( \sum_{n,m} \{ ^nL_{\pm}, {}^mL_{\mp} \} \right) \in \mathscr{G}_0$ iff $n - m = $ a constant.

The Spinor Superalgebra $S^2(N/2)$ is a superalgebra, as defined above, with $\dim(\mathscr{G}_0) = 3$ and $\dim(\mathscr{G}_1^{\pm}) = N$, where $N$ is odd. If we write the $2N+3$ basis elements of $S^2(N/2)$ as $\left\{ L_0, L_3, \hat{L} \right\} \in \mathscr{G}_0$ and $\left\{ ^nL_{\pm} \right\}_{n=1}^{N} \in \mathscr{G}_1^{\pm}$, then the following relations define $S^2(N/2)$:

$$[L_0, L_3] = 0 \,, \ [L_0, \hat{L}] = 0 \,, \ [L_3, \hat{L}] = 0 \tag{4.3a}$$

$$[L_0, {}^nL_{\pm}] = \pm {}^nL_{\pm} \,, \ [L_3, {}^nL_{\pm}] = \pm n \, {}^nL_{\pm} \,, \ [\hat{L}, {}^nL_{\pm}] = 0 \tag{4.3b}$$

$$\left\{ {}^nL_{\pm}, {}^mL_{\pm} \right\} = 0 \,, \ \sum_{n=1}^{N} \left\{ {}^nL_+, {}^nL_- \right\} = \hat{L} \tag{4.3c}$$

$$\sum_{n=1}^{N-m} \left\{ {}^nL_{\pm}, {}^{n+m}L_{\mp} \right\} = 0 \,, \text{ for } m = 1,2,..,N-1. \tag{4.3d}$$

The following is a realization of the elements of $S^2(N/2)$ in terms of $(N+1) \times (N+1)$ matrices of differential operators of degrees 0, 1 or 2. It lends itself to the interpretation of operators acting in an $N+1$ dimensional space that could support a representation of the Lorentz space-time symmetry group for spin $N/2$:

$$L_0 = \frac{1}{2}\left(\begin{array}{c|c} 1 & 0 \\ \hline 0 & -1 \end{array}\right), \quad L_3 = \frac{1}{2}\left(\begin{array}{c|c} \begin{smallmatrix} N & & \\ & N-2 & \quad 0 \\ & 0 & \ddots \\ & & & 3 \\ & & & & 1 \end{smallmatrix} & 0 \\ \hline 0 & \begin{smallmatrix} -1 & & \\ & -3 & \quad 0 \\ & 0 & \ddots \\ & & -N+2 \\ & & & -N \end{smallmatrix} \end{array}\right), \qquad (4.4a)$$

where $1$ is the $\frac{N+1}{2} \times \frac{N+1}{2}$ unit matrix. Additionally, we take

$$^{n}L_{\pm} = {}^{n}\sigma_{\pm}\left[W(x) \mp \frac{d}{dx}\right], \qquad (4.4b)$$

where,

$$^{1}\sigma_{+} = \left(\begin{array}{c|c} 0 & \begin{smallmatrix} 0 \\ \;\;0 \\ \times\;\; 0 \end{smallmatrix} \\ \hline 0 & 0 \end{array}\right), \quad ^{2}\sigma_{+} = \left(\begin{array}{c|c} 0 & \begin{smallmatrix} 0\;\; 0 \\ \times\;\; 0 \\ 0\;\times\;\; 0 \end{smallmatrix} \\ \hline 0 & 0 \end{array}\right), \quad ^{3}\sigma_{+} = \left(\begin{array}{c|c} 0 & \begin{smallmatrix} 0\;\; 0 \\ 0\;\times\;\; 0 \\ 0\;\times\;\; 0 \\ 0\;0\;\times\;\; 0 \end{smallmatrix} \\ \hline 0 & 0 \end{array}\right), \quad \text{etc.,} \quad ^{n}\sigma_{-} =$$

$\left({}^{n}\sigma_{+}\right)^{\dagger}$, and $\times$ stands for an arbitrary complex parameter. $W(x)$ is a real differentiable function of the configuration coordinate $x$ that contains the dynamics. The realization (4.4b) gives $\left({}^{n}L_{\pm}\right)^{\dagger} = {}^{n}L_{\mp}$, which implies, with the help of (4.3b) and (4.3c), hermiticity of the even operators. The number of complex parameters in $\left\{{}^{n}\sigma_{\pm}\right\}_{n=1}^{N}$ is $\left(\frac{N+1}{2}\right)^2$. However, we shall see that relation (4.3d) leaves only $\frac{N+1}{2}$ of them independent. The realization of the elements of $\mathscr{G}_1^{\pm}$ given by Eq. (4.4b) results in the following trivial relation:

$$\left\{{}^{n}L_{\pm}, {}^{n+m}L_{\mp}\right\} \equiv 0, \text{ for } N-n \geq m \geq \frac{N+1}{2}. \qquad (4.5)$$

Consequently, the only non-trivial set of relations in (4.3d) is the following:

$$\sum_{n=1}^{\frac{N+1}{2}}\left\{{}^{n}L_{\pm}, {}^{n+m}L_{\mp}\right\} = 0, \text{ for } m = 1, 2, .., \frac{N-1}{2}, \qquad (4.3d')$$

which gives $\frac{N^2-1}{4}$ constraint equations on the complex parameters of $\left\{{}^{n}\sigma_{\pm}\right\}_{n=1}^{N}$. Thus, the number of arbitrary parameters equals $\left(\frac{N+1}{2}\right)^2 - \frac{N^2-1}{4} = \frac{N+1}{2}$. Using this realization of $^{n}L_{\pm}$, we obtain the following from (4.3c):

$$\hat{L} \sim \left(\begin{array}{c|c} 1 \otimes \left(-\frac{d^2}{dx^2} + W^2 - W'\right) & 0 \\ \hline 0 & 1 \otimes \left(-\frac{d^2}{dx^2} + W^2 + W'\right) \end{array}\right). \qquad (4.4c)$$

This even operator forms the center of the superalgebra because it commutes with all of its elements. Moreover, since the subalgebra $\mathscr{G}_0$ that contains $\hat{L}$ is abelian and of dimension three, then all representations of $S^2(N/2)$ could be parameterized by three real numbers corresponding to the eigenvalues of each of these three, mutually commuting, hermitian operators. We refer to such a representation by $D_{\pm}^{N}(E,k)$, where $E$ is a real parameter and $k = 0,1,..,\frac{N-1}{2}$. If the basis elements of this representation are labeled $\left\{ \phi_k^{\pm}(x,E) \right\}$, then we could write the following:

$$L_0 \left| \phi_k^{\pm} \right\rangle = \pm \left| \phi_k^{\pm} \right\rangle, \; L_3 \left| \phi_k^{\pm} \right\rangle = \pm \left( k + \frac{1}{2} \right) \left| \phi_k^{\pm} \right\rangle, \tag{3.9a}$$

$$\hat{L} \left| \phi_k^{\pm} \right\rangle \sim \left( -\frac{d^2}{dx^2} + W^2 \mp W' \right) \left| \phi_k^{\pm} \right\rangle = 2E \left| \phi_k^{\pm} \right\rangle. \tag{3.9b}$$

Consequently, we may interpret $L_3$ as the spin projection operator and $L_0$ as the parity operator. Moreover, in supersymmetric quantum mechanics [24], $W^2 \pm W'$ are the two isospectral (except for the ground state) superpotentials giving an interpretation to $\hat{L}$ as the associated Hamiltonian and $E$, the energy. We can write a linear hermitian operator $Q \in S^2(N/2)$ as:

$$Q = \lambda_0 L_0 + \lambda_3 L_3 + \hat{\lambda}\hat{L} + \sum_{n=1}^{N} \left( \lambda_n^+ \, {}^n L_+ + \lambda_n^- \, {}^n L_- \right), \tag{4.6}$$

where $\left\{ \lambda_0, \lambda_3, \hat{\lambda} \right\}$ are real parameters and $\lambda_n^- = \left( \lambda_n^+ \right)^*$. If we choose the representation $D_{\pm}^{N}(E,k)$, then compatibility with Eq. (3.9b) requires that $\lambda_3 = 0$. Additionally, if $Q$ is to be first order in the derivatives, then $\hat{\lambda} = 0$.

## V. DISCUSSION

In three dimensions, the spinor wave function $\chi = \begin{pmatrix} \phi^+ \\ \phi^- \end{pmatrix}$ for a spin $1/2$ particle in Eq. (2.3) has four components, with $\phi^{\pm}(\vec{r})$ each being a two-component element. For a spherically symmetric interaction, the Hamiltonian commutes with the total angular momentum $\vec{\mathcal{J}} = \vec{\mathcal{L}} + \vec{\mathcal{S}}$, but neither separately with the orbital component $\vec{\mathcal{L}}$ or the spin component $\vec{\mathcal{S}}$. Therefore, $\chi$ is a simultaneous eigenfunction of $H$, $\mathcal{J}^2$, and $\mathcal{J}_3$. However, the two component spinors $\phi^{\pm}$ are, additionally, eigenfunctions of $\mathcal{L}^2$ and $\mathcal{S}^2$. Thus, they should also be eigenfunctions of $\vec{\mathcal{S}} \cdot \vec{\mathcal{L}}$. Therefore, $\phi^{\pm}$ should satisfy the following relations:

$$\mathcal{J}^2 \phi^{\pm} = j(j+1)\phi^{\pm}, \; \mathcal{L}^2 \phi^{\pm} = \ell(\ell+1)\phi^{\pm}, \; \mathcal{S}^2 \phi^{\pm} = \tfrac{3}{4}\phi^{\pm}, \; \mathcal{J}_3 \phi^{\pm} = m\phi^{\pm}, \tag{5.1}$$

where $j = \left| \ell - \frac{1}{2} \right|$ or $j = \ell + \frac{1}{2}$, and for which $m = -j, -j+1,.., j$. On the other hand, we should also have the following relation:

$$2\vec{\mathcal{S}} \cdot \vec{\mathcal{L}} \phi^{\pm} = \left[ j(j+1) - \ell(\ell+1) - \tfrac{3}{4} \right] \phi^{\pm} \equiv -(\kappa+1)\phi^{\pm}, \tag{5.2}$$

where the kinematic constant $\kappa$ takes the following values:

$$\kappa = \begin{cases} \ell = 1,2,3,.. & , j = \ell - \frac{1}{2} \\ -\ell - 1 = -1,-2,-3,.. & , j = \ell + \frac{1}{2} \end{cases} \tag{5.3}$$

In such 3D problems with spherical symmetry, the radial Dirac Hamiltonian is obtained from that in Eq. (2.2) by the replacement $W(x) \rightarrow W(r) + \frac{\kappa}{r}$ [13,15,18]. Therefore, the canonical problem (where, $V = S = 0$) results in the following wave equation:

$$\left[-\frac{d^2}{dr^2}+\frac{\kappa(\kappa\pm1)}{r^2}+W^2\mp W'+2\kappa\frac{W}{r}-\frac{\varepsilon^2-1}{\hbar^2}\right]\varphi^\pm(r)=0\,,\qquad(5.4)$$

where the radial component of $\phi^\pm(\vec{r})$ is $\frac{1}{r}\varphi^\pm(r)$. This equation is to be compared with Eq. (2.10) for the one dimensional problem.

For a spin $3/2$ particle in three spatial dimensions, the wave function $\psi$ in Eq. (3.1) is an eight-component spinor with $\left\{\phi_k^\pm(\vec{r})\right\}_{k=0,1}$ having two components each. In the case of spherical symmetry, these components satisfy the same set of equations in (5.1), except for the third one, which should read $\mathcal{S}^2\phi_k^\pm=\frac{15}{4}\phi_k^\pm$, and where $j=\left|\ell-\frac{3}{2}\right|$, $\left|\ell-\frac{1}{2}\right|$, $\ell+\frac{1}{2}$, $\ell+\frac{3}{2}$. Moreover, Eq. (5.2) should be replaced by

$$2\vec{\mathcal{S}}\cdot\vec{\mathcal{L}}\phi_k^\pm=\left[j(j+1)-\ell(\ell+1)-\frac{15}{4}\right]\phi_k^\pm\equiv-(\tau+1)\phi_k^\pm\,,\qquad(5.5)$$

where the constant $\tau$ takes the following values:

$$\tau=\begin{cases}3\ell+2=8,11,14,... & ,j=\ell-\frac{3}{2}\\ \ell+3=4,5,6,... & ,j=\ell-\frac{1}{2}\\ -\ell+2=2,1,0,... & ,j=\ell+\frac{1}{2}\\ -3\ell-1=-1,-4,-7,... & ,j=\ell+\frac{3}{2}\end{cases}\qquad(5.6)$$

In analogy with spin $1/2$, the radial Dirac Hamiltonian is obtained from that in Eq. (3.20) by the replacement: $W(x)\rightarrow W(r)+\frac{\tau}{r}$. Therefore, the canonical problem (where, $V=0$) results in the following wave equation for the radial spinor components:

$$\left[-\frac{d^2}{dr^2}+\frac{\tau(\tau\pm1)}{r^2}+W^2\mp W'+2\tau\frac{W}{r}-\frac{\varepsilon^2-1}{\hbar^2}\right]\varphi_k^\pm(r)=0\,.\qquad(5.7)$$

As an example, we calculate the relativistic energy spectrum of the Dirac-Oscillator problem for spin $3/2$. In this settings, the Dirac-Oscillator [26] is defined with $V=0$ and $W=\omega^2 r$, where $\omega$ is the oscillator frequency. Substituting these in Eq. (5.7) gives the following Schrödinger-type wave equation for each of the four spinor components:

$$\left[-\frac{d^2}{dr^2}+\frac{\tau(\tau\pm1)}{r^2}+\omega^4 r^2+\omega^2\left(2\tau\mp1\right)-\frac{\varepsilon^2-1}{\hbar^2}\right]\varphi_k^\pm(r)=0\,,\qquad(5.8)$$

where $k=0$ or 1. We Compare this with the nonrelativistic wave equation for the three dimensional isotropic oscillator:

$$\left[-\frac{d^2}{dr^2}+\frac{\mathscr{L}(\mathscr{L}+1)}{r^2}+\omega^4 r^2-2E\right]\Phi(r)=0\,,\qquad(5.9)$$

where $\mathscr{L}$ is the nonrelativistic angular momentum quantum number, and $E$ is the energy. The comparison gives the following two maps between the relativistic and non-relativistic problems:

for $\varphi_k^+$: $E\rightarrow\frac{\varepsilon^2-1}{2\hbar^2}-\omega^2\left(\tau-\frac{1}{2}\right)$, $\mathscr{L}\rightarrow\begin{cases}\tau\\-\tau-1\end{cases}$ $\qquad(5.10a)$

for $\varphi_k^-$: $E\rightarrow\frac{\varepsilon^2-1}{2\hbar^2}-\omega^2\left(\tau+\frac{1}{2}\right)$, $\mathscr{L}\rightarrow\begin{cases}\tau-1\\-\tau\end{cases}$ $\qquad(5.10b)$

The top (bottom) choice of the $\mathscr{L}$ map corresponds to $\ell>j$ ($\ell<j$). Using these parameter maps and the well-known nonrelativistic energy spectrum, $E_n=\omega^2\left(2n+\mathscr{L}+3/2\right)$, we obtain the following relativistic spectrum:

$$\left| \varepsilon_n^{\pm} \right| = \begin{cases} \sqrt{1 + 4\hbar^2 \omega^2 \left( n + \tau + 1/2 \right)} & , \quad \ell > j \\ \sqrt{1 + 4\hbar^2 \omega^2 \left( n + \frac{1 \mp 1}{2} \right)} & , \quad \ell < j \end{cases} \tag{5.11}$$

where $n = 0, 1, 2, \dots$. Using Eq. (5.6) for $\tau$, this could explicitly be written as:

$$\left| \varepsilon_n^{\pm} \right| = \begin{cases} \sqrt{1 + 4\hbar^2 \omega^2 \left( n + 3\ell + 5/2 \right)} & , j = \ell - \frac{3}{2} \\ \sqrt{1 + 4\hbar^2 \omega^2 \left( n + \ell + 7/2 \right)} & , j = \ell - \frac{1}{2} \\ \sqrt{1 + 4\hbar^2 \omega^2 \left( n + \frac{1 \mp 1}{2} \right)} & , j = \ell + \frac{1}{2}, j = \ell + \frac{3}{2} \end{cases} \tag{5.11'}$$

Therefore, the lowest (highest) positive (negative) energy state, where $\varepsilon = +1$ ($\varepsilon = -1$), occurs for $\ell < j$. It is associated with the spinor wavefunction $\psi = \left( \phi_1^+, \phi_0^+, 0, 0 \right)_{n=0}^{\top}$. This is the only non-degenerate state while all others are degenerate because the energy spectrum in (5.11) satisfies:

$$\begin{aligned} \varepsilon_n^- &= \varepsilon_n^+ &, \ell > j \\ \varepsilon_n^- &= \varepsilon_{n+1}^+ &, \ell < j \end{aligned} \tag{5.12}$$

It is also worthwhile noting that the radial components of the spinor wavefunction could be obtained by using the same parameter maps (5.10) and the nonrelativistic wavefunction $\Phi_n(r) \sim (\omega r)^{\ell+1} e^{-\omega^2 r^2 / 2} L_n^{\ell+1/2}(\omega^2 r^2)$, where $L_n^\mu(z)$ are the orthogonal Laguerre polynomials.